\documentclass[graybox]{svmult}

\usepackage{mathptmx}       
\usepackage{helvet}         
\usepackage{courier}        
\usepackage{type1cm}        
\usepackage{makeidx}         
\usepackage{graphicx}        
\usepackage{multicol}        
\usepackage[bottom]{footmisc}
\usepackage{amsfonts}

\makeindex             

\newcommand{\beq}{\begin{equation}}
\newcommand{\eeq}{\end{equation}}
\newcommand{\beqa}{\begin{eqnarray}}
\newcommand{\eeqa}{\end{eqnarray}}
\newcommand{\nn}{\nonumber \\}

\def \e {\mathrm{e}}

\def \la {\langle}
\def \ra {\rangle}
\def \s {\sigma}
\def \t {\tau}

\def \B {{\mathcal B}}
\def \C {{\mathbb C}}

\def \H {{\mathcal H}}

\def \ch {\mathrm{ch}}

\def \el {\mathrm{el}}

\def \z {\zeta}

\def \Id {{\mathbb I}}

\def \mod {\ \mathrm{mod} \ }
\def \H {{\mathcal H}}
\def \uu {{\widehat{u(1)}}}
\def \P {{\mathcal P}}
\begin{document}
\title*{Topological Quantum Computation with non-Abelian  anyons in fractional quantum Hall states}
\titlerunning{TQC with non-Abelian  anyons} 
\author{Lachezar S. Georgiev}
\institute{Lachezar S. Georgiev \at Institute for Nuclear Research and Nuclear Energy, Bulgarian Academy of Sciences, 
72 Tsarigradsko Chaussee, Sofia, Bulgaria \email{lgeorg@inrne.bas.bg}
}
%
%
\maketitle
\abstract{
We review the general strategy of topologically protected quantum information processing based on non-Abelian anyons, in which 
quantum information is encoded into the fusion channels of pairs of anyons and in fusion paths for multi-anyon states, 
realized in two-dimensional fractional quantum Hall systems. The quantum gates which are needed for the quantum information processing 
 in these multi-qubit registers are implemented by exchange or braiding of the non-Abelian anyons that are at fixed positions in 
two-dimensional coordinate space.
As an example we consider the Pfaffian topological quantum computer based on the fractional quantum Hall state with filling factor $\nu_H=5/2$.
The elementary qubits are constructed by localizing Ising anyons on fractional quantum Hall antidots and  various quantum gates, 
such as the Hadamard gate, phase gates and CNOT, are explicitly realized by braiding. We also discuss the appropriate experimental 
signatures which could eventually be used to detect non-Abelian anyons in Coulomb blockaded quantum Hall islands.
}
\section{Introduction: Quantum Computation in general}
\label{sec:QC}
Quantum Computation (QC) is a relatively new field \cite{nielsen-chuang} of computational research in which information is encoded in 
two-level quantum systems called \textit{qubits}, or \textit{qu}antum \textit{bits}, in analogy with the classical bits. In contrast with the classical bits,  
which can have only two states `0' and `1', the qubit is
an arbitrary linear combination of the two basic quantum states $|0\ra$ and $|1\ra$
\beq \label{qubit}
|\psi\ra = \alpha|0\ra +\beta |1\ra, \quad \mathrm{with} \quad \alpha, \beta \in \C \quad \mathrm{and} \quad 
 |\alpha|^2+|\beta|^2=1 ,
\eeq
i.e., a qubit $|\psi\ra$ belongs to the projective $\C^2$ space which is called the Bloch sphere \cite{nielsen-chuang}.
In order to prepare one qubit state for QC we have to  define first an orthonormal basis $|0\ra$,  $|1\ra$, which could be for example
the two spin-projection states of a spin 1/2 particle, and then  construct physically the state $|\psi\ra= |0\ra$ or  $|\psi\ra= |1\ra$.

One fundamental aspect of QC is the measurement procedure, which is described mathematically by a 
collection $\{M_m\}$ of measurement operators representing the $m$-th outcome of the measurement
and satisfy the completeness condition
\[
\sum_m M_m^\dagger M_m =\Id .
\]
In more detail, if the state before measurement is $|\psi \ra $  then the probability for outcome $m$ in the measurement  of the 
observable $M$ is
\[
p(m)=\la  \psi | M_m^\dagger M_m  | \psi \ra ,
\]
 and the state after measurement is a projection $|\psi ' \ra$ of the initial state
\[
|\psi ' \ra =\frac{M_m|\psi\ra}{\sqrt{\la \psi |   M_m^\dagger M_m |\psi\ra}} .
\]
Another important condition for QC is that the  states in the computational basis must be orthonormal,
otherwise no measurement can distinguish between them.
Next, just like in classical computation, where we need a register of bits  which belong to the direct sums of bit spaces, 
QC requires the construction of a register of qubits, or, multiple-qubit, which however belong to the  tensor product of the single-qubit 
spaces, e.g., the $n$-qubit register belongs to the space 
\[
\H^n = \underbrace{\C^2\otimes \C^2 \cdots \C^2}_n   \simeq \left(\C^{2}\right)^n   \simeq \C^{2^n}  , 
\]
that should be projective by definition, thus expressing  the required normalization of the $n$-qubit states \cite{nielsen-chuang}.
This difference in the register spaces dimension in the classical (direct sum space has dimension $2n$) and quantum computation 
(tensor product space has dimension $2^n$) is one of the 
reasons for the significantly bigger computational power of the quantum computers as compared to the classical ones \cite{nielsen-chuang}.

The first step in a concrete QC is the initialization of the $n$-qubit register, which is done in analogy with the initialization of a classical one, i.e., 
\[
010011000...01 \quad \mathop{\to}\limits^{\mathrm{QC}} \quad |010011000...01\ra ,
\]
where we used the standard notation for the tensor-product basis states, e.g., $|01\ra = |0\ra\otimes |1\ra$ and 
$|101\ra = |1\ra \otimes |0\ra\otimes |1\ra$, etc.

The second step in QC is to process the initial information, which is done by applying a quantum operation, called \textit{quantum gate},
 on the $n$-qubit register \cite{nielsen-chuang}. The quantum gates are realized by unitary quantum operators acting over the space $\H^n$. 
Sequential application of two quantum gates is a quantum gate again and any quantum gate possesses its inverse. 
In other words, quantum gates $G$  should realize by appropriate physical processes all unitary $2^n \times 2^n$ matrices and therefore 
belong to the unitary group $G\in SU(2^n)$. 

The third step in QC is the measurement of a given observable, as described above, of the $n$-qubit register after its processing 
with a number of quantum gates and this measurement is the result of the QC.
\begin{svgraybox}
It is important to remember, that universal quantum computation requires physical realization of all unitary matrices in the unitary group 
$SU(2^n)$ for an arbitrary finite number $n$ of qubits and their measurements. 
\end{svgraybox}
After the work of Peter Shor \cite{shor}, unveiling an algorithm for factorization of large numbers $N$ into primes on a quantum computer, 
which requires time that is polynomial in the size $\log N$, while classical factorization algorithms require exponential time, it has become 
obvious that quantum computers, if they can be constructed, would be much faster than classical ones. This anticipated 
exponential speed-up with respect to  classical computing is due to the fundamental properties of the quantum systems, such as  
quantum parallelism and  entanglement \cite{nielsen-chuang}. The difficulties to factorize large numbers on classical computers is the 
security foundation 
of some public-key cryptographic algorithms, such as the RSA, which explains the wide interest of the banks, intelligence services 
 and military services   in the physical realization of quantum computers.

Another important theoretical observation in the field of QC is that any unitary operator, i.e., any quantum gate,
 can be approximated by products (sequential application, or concatenation) \cite{nielsen-chuang} of only 3 universal gates: 
$H$,  $T$ and CNOT, that can be applied to any qubit in the $n$-qubit register,
with arbitrary precision in the distance
\[
E(U,V)\equiv \max_{|\psi\ra} || \left(U-V \right)|\psi\ra|| .
\]
The first two quantum gates, $H$ and $T$ act on single qubits, while CNOT is a quantum gate acting on two qubits in the $n$-qubit register.
These three basic quantum gates can be written explicitly, in the basis $\{|0\ra, |1\ra \}$ for $H$ and $T$ and  in the two-qubit basis
$  \{|00\ra, |01\ra,  |10\ra, |11\ra\} $ for CNOT as \cite{nielsen-chuang}
\begin{svgraybox}
\beq \label{discrete}
H=\frac{1}{\sqrt{2}}\left[ \matrix{1 & \ \ \  1 \cr 1 & -1}\right], \quad
T= \left[ \matrix{1 &  0 \cr 0 & \e^{i\pi/4}}\right], \quad 
\mathrm{CNOT}=
\left[ \matrix{1 & 0 & 0 & 0 \cr 0 & 1 & 0 & 0 \cr  0 & 0 & 0 &  1 \cr  0 & 0 & 1 & 0}\right] .
\eeq
\end{svgraybox}
This observation significantly simplifies the task of physical implementation of quantum computers, reducing the 
variety of different type of quantum operations, which have to be  constructed by different physical processes, to only three simple gates. 
However, there are huge difficulties on the way of constructing a stable quantum computer due to the unavoidable decoherence and 
noise \cite{nielsen-chuang}  resulting from the local interactions of the qubits with their environment,  destroying in this way all 
coherent phenomena and flipping uncontrollably $|0\ra \leftrightarrow |1\ra$.

One way to make the fragile quantum information more robust is to use the so-called quantum error-correcting algorithms \cite{nielsen-chuang}.
This brings in a new hope for the physical realization of the quantum computers by compensating hardware deficiency through
 clever circuit design and increasing the number of qubits and quantum gates,  so that even if some information is corrupted it could be 
restored at the end. The bad news is that all quantum error-correcting algorithms require big overhead in the number of qubits and quantum 
gates.

Another possible way out might be the topological quantum computation (TQC) \cite{kitaev-TQC,sarma-RMP}, whose strategy is to 
improve QC hardware by using intrinsic topological protection  instead of compensating hardware deficiency by clever circuit design. 
This new idea requires a fundamentally new concept: 
the non-Abelian exchange  statistics of quasiparticles which are believed to exist in the fractional quantum Hall states \cite{sarma-RMP}.
\section{Non-Abelian anyons and topological QC}
\label{sec:NA}
\begin{figure}[htb]
\centering
\includegraphics[bb=40 490 540 630,clip,width=12cm]{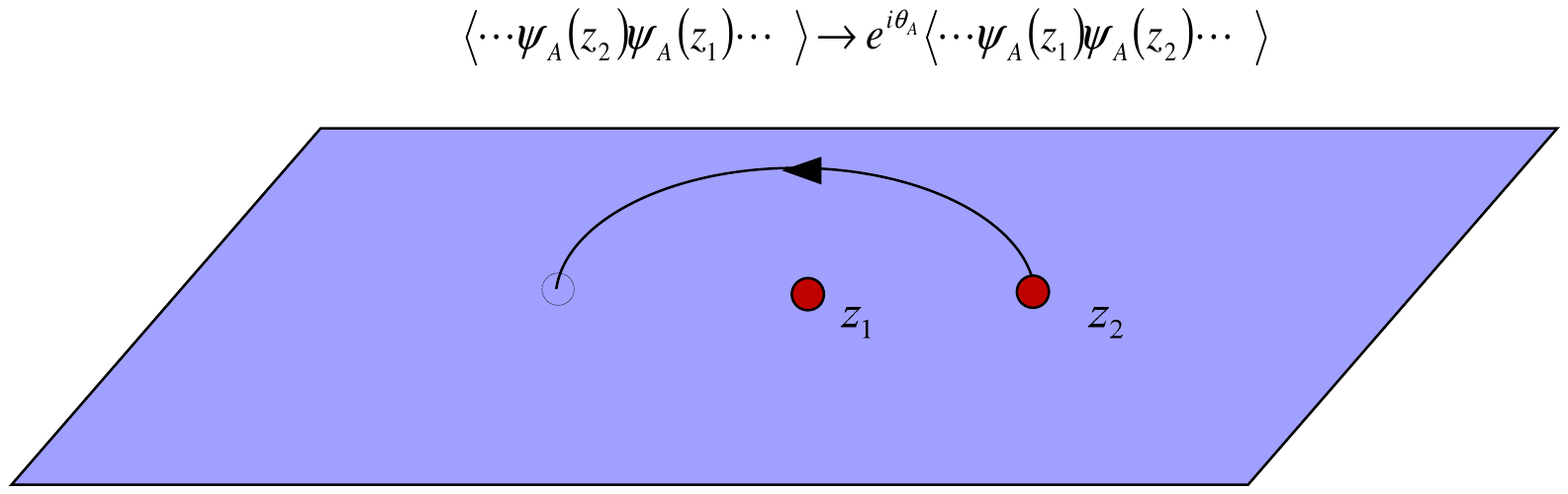}
\caption{Adiabatic counter-clockwise exchange of two identical particles in the complex plane \label{fig:stat-2}}
\end{figure}

In this section we will explain the new concept of non-Abelian exchange statistics of particles and will try to give an idea of what it 
can be used for. Consider a system of indistinguishable particles at fixed positions in space. 
One of their important statistical characteristics is their statistical angle $\theta_A/\pi$, which can be
defined in the following way: the quantum state of the system with many indistinguishable particles of type `A'
can be expressed as a correlation function (or a vacuum expectation value) of some quantum filed operators 
$\psi_A(z)$ which represent the act of creation of a particle of type `A' at position $z$ in the space. Let us focus only on two particles
of type `A', at positions $z_1$ and $z_2$ in space, although the system may contain  more particles. If we exchange the two 
particles adiabatically, as shown in Fig.~\ref{fig:stat-2}, then the state after the exchange differs from that before it  by the statistical  
phase  $\e^{i \pi (\theta_{A}/\pi)}$, i.e.,
\[
\la \cdots \psi_A(z_1)\psi_A(z_2) \cdots \ra \to  \la \cdots \psi_A(z_2)\psi_A(z_1) \cdots \ra  = 
 \e^{i \pi (\theta_{A}/\pi)}  \la \cdots \psi_A(z_1)\psi_A(z_2) \cdots \ra ,   
\]
where $\theta_{A}/\pi$ is by definition the statistical angle of the particles of  type `A'.
In three-dimensional space (or four-dimensional space-time) only two type of particles could exist, as long as the statistical angle 
is concerned\footnote{for simplicity we disregard here the case of relativistic parafermions and parabosons}: 
bosons, which correspond to $\theta_A/\pi=0$ and  fermions which correspond to $\theta_A/\pi=1$ .
However, in two-dimensional space the variety of statistical angles is much richer than in three dimensions.
For example, the so-called Laughlin anyons  correspond to $\theta_{L}/\pi=1/3$, and actually all values of 
$\theta_{A}/\pi$ between 0 and 1 are admissible. That is why such particles, which could exist only in two-dimensional 
(and one-dimensional) space, are called any-ons, in analogy with the bosons and fermions, as compared in Table~\ref{tab:anyons}.
\begin{table}[htb]
\caption{Statistical angles for bosons, anyons and fermions \label{tab:anyons}}
\centering
\begin{tabular}{c|c|c}
\hline
``bos-ons'' & ``any-ons'' & ``fermi-ons'' \cr\hline\hline
 & & \cr
\mbox{\qquad} $\theta_A/\pi=0$ \mbox{\qquad} & \mbox{\qquad}  $0 < \theta_A/\pi < 1$ \mbox{\qquad}  & \mbox{\qquad}  $\theta_A/\pi=0$ \mbox{\qquad}  \cr
& & \cr
\hline
\end{tabular}
\end{table}
The mathematical reason for this distinction is that the rotation group
$SO(3)$ of the  three dimensional space has a compact simply connected covering group $SU(2)$ and therefore 
any $4\pi$-rotation is equivalent to $\Id$.  Because of this, any $2\pi$-rotation in the three-dimensional space 
is  $U(2\pi)=\e^{-2\pi i J} =\pm \Id$ and this implies that $\theta/\pi= 2J =0$ or $1 \mod 2$, where $J$ is the eigenvalue of the 
spin and this expresses the well-known spin--statistics relation \cite{thouless:top,fro-stu-thi}.
On the other hand, the group $SO(2)$ of the rotations in the two-dimensional space does not have a compact covering group 
and therefore \textit{there is no boson--fermion alternative in two dimensions}, which allows for the statistical angle $\theta/\pi$
to be \textit{any real number between 0 and 1}.
\subsection{Construction of $n$-particle states: the braid group}
Another important distinction between the three-dimensional coordinate space and the two-dimensional one can be seen
in the many-particle states. The many-particle quantum states in three-dimensional space are built as representations of the 
symmetric group $\mathcal{S}_n$, which are symmetric for bosons and  antisymmetric for fermions. The  symmetric
group $\mathcal{S}_n$, which is the group of permutation of $n$ objects, is a finite group generated by $n-1$ elementary
transpositions of neighboring objects i.e., $\s_i: i \leftrightarrow i+1$ and obviously satisfy $\s_i=\s_i^{-1}$. Like any finite group 
$\mathcal{S}_n$ can be defined by its genetic code, i.e., the relations between its generators \cite{wilson}
\[
	\s_i \s_j = \s_j \s_i, \quad |i-j| \geq 2, 
\]
which means that the disjoint transpositions commute with each other, and
\[
	\s_i \ \s_{i+1} \ \s_i=\s_{i+1} \  \s_i \  \s_{i+1} , \quad   \left(\s_i \right)^2=\Id.
\]
On the other hand the many-particle quantum states in two dimensional  space are constructed as representations of the 
braid group \cite{birman} $\B_n$.
The braid group  $\B_n$ is an infinite group, which is an extension of the symmetric group $\mathcal{S}_n$ whose generators, 
unlike those for $\mathcal{S}_n$, do not satisfy $(B_i)^2=\Id $ .
The genetic code of the braid group  $\B_n$ is know as the \textit{Artin relations} \cite{birman}
\beqa
B_i B_j &=&  B_j B_i, \qquad \qquad \mathrm{for}  \quad |i-j|\geq 2 \nn
B_i B_{i+1} B_i &=&  B_{i+1} B_i B_{i+1}, \quad i=1,\ldots , n-1 .\nonumber
\eeqa
The very concept of the non-Abelian anyons requires the existence of degenerate multiplets  of $n$-particle states, 
with fixed coordinate positions of the anyons, and
the exchanges of the coordinates of these anyons generate statistical phases  $\e^{i\theta_A}$ which might be non-trivial matrices  
 acting on those multiplets. That is why this statistics is called non-Abelian   (see the example below).
 \subsection{Fusion paths: labeling anyonic states of matter}
As mentioned above, the states with many non-Abelian anyons at fixed coordinate positions form degenerate multiplets 
which means that specifying the positions of the anyons and their quantum numbers, such as the electric charge, single-particle 
energies and angular momenta, are not sufficient to specify a concrete $n$-particle state. 
More information is needed and this information is a non-local characteristics of the $n$-particle state as a whole.

This additional information is provided by the so-called \textit{fusion channels} \cite{CFT-book}. 
Consider the fusion of two particles of type `a' and `b',
i.e., consider the process when the two particles come to the same coordinate position and form a new particle of type `c'.
This can be written formally as
\[
\Psi_a \times   \Psi_b =   \sum\limits_{c=1}^{g}   N_{ab}{}^c   \Psi_c ,
\]
where the fusion coefficients $(N_{ab})^c$ are integers, specifying the different possible channels of the fusion process, which 
are symmetric and associative \cite{CFT-book}. Two particles  $\Psi_a$ and $\Psi_b$, where $b$ could 
be the same as $a$, are called non-Abelian anyons if the fusion coefficients $N_{ab}{}^c \neq 0$ for more than one $c$.  
The most popular example of non-Abelian anyons are the Ising anyons realized in two-dimensional conformal field theory
with $\uu \times \mathrm{Ising}$ symmetry by the primary field \cite{CFT-book}
\[
\Psi_I(z)=\s(z)\ :\e^{i\frac{1}{2\sqrt{2}}\phi(z)}: ,
\]
where $\phi(z)$ is a normalized $\uu$ boson \cite{CFT-book} and $\s(z)$ is the chiral spin field of the two-dimensional Ising model 
CFT \cite{CFT-book}.
The  fusion rules of the Ising model are non-Abelian since the fusion process of two Ising anyons $\s$
could be realized in two different fusion channels: that of the identity operator $\Id$ and that of the Majorana 
fermion $\psi$ \cite{CFT-book}, i.e.,
\beq \label{fusion}
  \s \times \s = \Id +\psi, \quad \s \times \psi = \s .
\eeq
The quantum information is  then encoded  into the fusion channel and the computational basis is defined by pairs of Ising anyons
whose fusion channel is fixed
\beqa\label{encoding}
|0\ra=(\s,\s)_{\Id}\quad\ &\longleftrightarrow& \quad \s \times \s  \to \Id \nn
|1\ra=(\s,\s)_{\psi}\quad &\longleftrightarrow& \quad \s \times \s  \to \psi ,
\eeqa
i.e. the state is $|0\ra$ if the two Ising anyons $\s$ fuse to $\Id$ and $|1\ra$ if they fuse to $\psi$.
The fusion channel is a topological quantity--it is independent of the fusion process details and  depends only on the topology  
of the coordinate space with positions of the anyons removed. This quantity is non-local: the fusion channel is independent of 
the anyon separation and is preserved after separation. It is also robust and persistent: if we fuse two particles and then split them again,
 their fusion channel does not change.  

The concatenation of several fusion channels of neighboring pairs of anyons is called a \textit{fusion paths} and can be
displayed  in Bratteli diagrams \cite{sarma-RMP}, see Fig.~\ref{fig:bratteli} below.
\begin{figure}[htb]
\centering
\includegraphics*[bb=0 450 340 630,width=8.5cm]{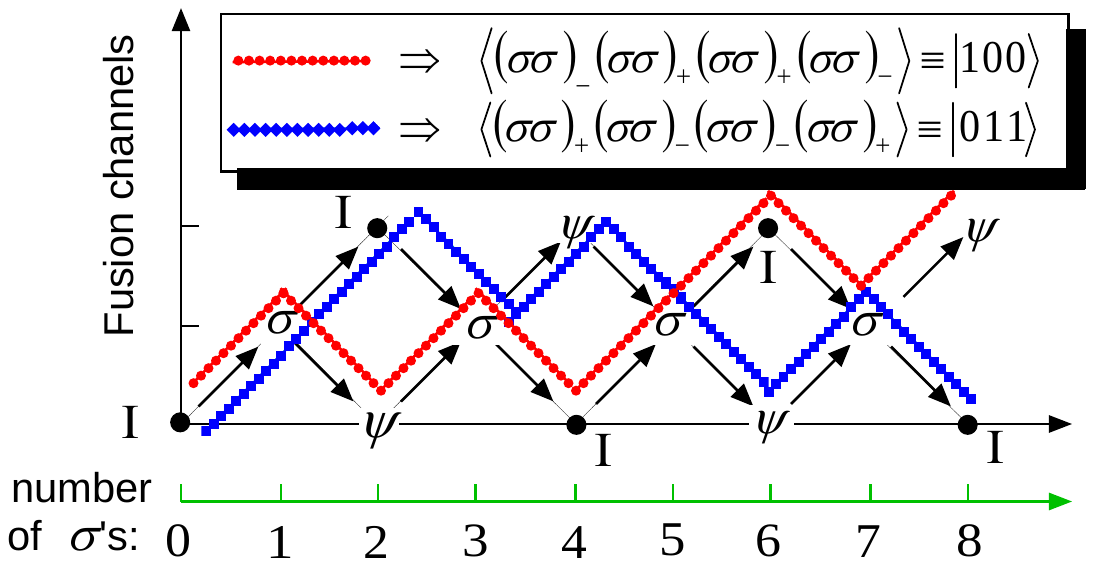} 
\caption{\textbf{Bratteli diagram for Ising anyons. Each step to the right on the horizontal axis represents fusion with one more 
Ising anyon field $\s$\label{fig:bratteli}}}
\end{figure}
The blue (square) and the red (dot) fusion paths represent two different quantum states,  $| 011\ra$ and $| 100\ra$ respectively, of 8 Ising anyons 
at fixed positions (cf. Ref.~\cite{clifford}), which belong to the same degenerate multiplet.
\begin{svgraybox}
In other words, quantum states with many non-Abelian anyons at fixed positions in two-dimensional space
 are specified/labeled by fusion paths and can  be plotted in Bratteli diagrams.
\end{svgraybox}
\subsection{Braiding of anyons: topologically protected quantum gates}
\begin{figure}[htb]
\centering
\includegraphics*[bb=5 380 590 790,width=\textwidth]{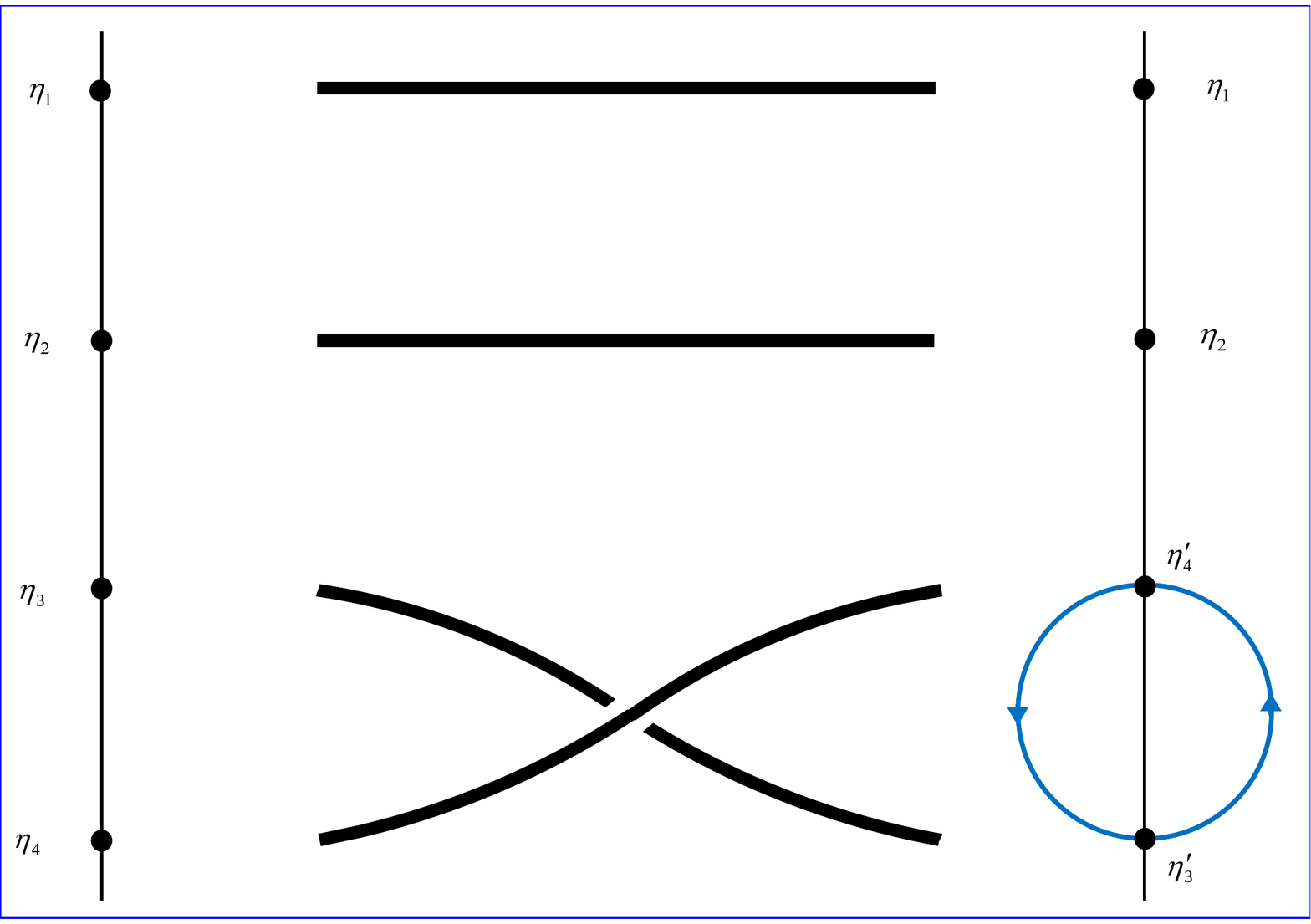}
\caption{Braiding of two anyons with coordinates $\eta_3$ and $\eta_4$ in a state containing 4 Ising anyons at fixed 
positions with coordinates $\eta_1$, \ldots, $\eta_4$. Time is running on the horizontal axis to the right.\label{fig:R_34}}
\end{figure}

Quantum operations in TQC, needed for processing of quantum information, are implemented by braiding non-Abelian 
anyons \cite{sarma-RMP,stern-review}.
Braiding of two anyons is the adiabatic exchange of the coordinate positions of the two anyons in the counter-clockwise 
direction, without crossing any other coordinates as illustrated in Fig.~\ref{fig:R_34}.
In this figure  we assumed that the time during the adiabatic exchange is on the horizontal axis, while the positions at any 
moment is on the vertical axis. The resulting diagram is called a braid diagram and can be used to represent graphically 
the braiding process.

\begin{figure}[htb]
\centering
\includegraphics*[bb=0 435 540 570,width=8cm]{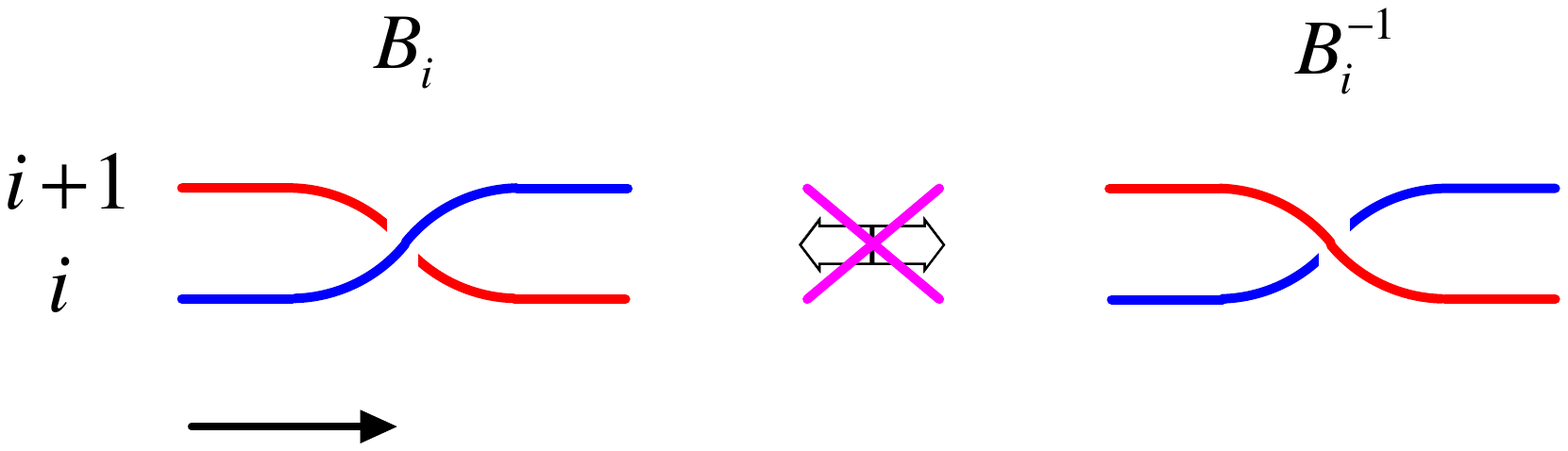}
\caption{Braid diagrams: exchange of particles with number $i$ and $i+1$ in counter-clockwise and clockwise 
 directions are distinct and inverse to each other. \label{fig:braids} }
\end{figure}

The clockwise exchanges represent the inverse of the braids described above and they are not equal to the 
counter-clockwise exchanges as illustrated in Fig.~\ref{fig:braids}.

For the purpose of illustration of the non-Abelian statistics acting on a degenerate multiplet of multyanyon states
we consider the wave functions representing 8 Ising anyons $\s$ at fixed positions with coordinates $\eta_1, \ldots, \eta_8$,
which correspond to three-qubit states. Each pair of Ising anyons $\s$ is characterized by the fermion parity of 
their fusion channel in Eq.~(\ref{fusion}), i.e., it is `$+$' for the fusion channel of the identity operator $\Id$ and 
`$-$' for the fusion channel of the Majorana fermion $\psi$. Taking into account the information encoding (\ref{encoding})
in the quantum information language we note that `$+$' corresponds to the computational state $|0\ra$ while `$-$' corresponds
 to $|1\ra$ and therefore the multiplet can be explicitly written in the three-qubit computational basis as
\beqa\label{3qubits}
|000\ra &\equiv& \la [\s(\eta_1) \s(\eta_2)]_+[\s(\eta_3)\s(\eta_4)]_+[\s(\eta_5)\s(\eta_6)]_+[\s(\eta_7)\s(\eta_8)]_+\ra \nn
|001\ra &\equiv& \la [\s(\eta_1) \s(\eta_2)]_+[\s(\eta_3) \s(\eta_4)]_+[\s(\eta_5)\s(\eta_6)]_-[\s(\eta_7)\s(\eta_8)]_-\ra \nn
|010\ra &\equiv& \la [\s(\eta_1)\s(\eta_2)]_+[\s(\eta_3) \s(\eta_4)]_-[\s(\eta_5)\s(\eta_6)]_+ [\s(\eta_7)\s(\eta_8)]_-\ra \nn
|011\ra &\equiv& \la [\s(\eta_1)\s(\eta_2)]_+[\s(\eta_3) \s(\eta_4)]_-[\s(\eta_5)\s(\eta_6)]_-[\s(\eta_7)\s(\eta_8)]_+\ra \nn
|100\ra &\equiv& \la [\s(\eta_1)\s(\eta_2)]_-[\s(\eta_3) \s(\eta_4)]_+[\s(\eta_5)\s(\eta_6)]_+[\s(\eta_7)\s(\eta_8)]_-\ra \nn
|101\ra &\equiv& \la [\s(\eta_1)\s(\eta_2)]_-[\s(\eta_3) \s(\eta_4)]_+[\s(\eta_5)\s(\eta_6)]_-[\s(\eta_7)\s(\eta_8)]_+\ra \nn
|110\ra &\equiv& \la [\s(\eta_1)\s(\eta_2)]_-(\s(\eta_3) \s(\eta_4)]_-[\s(\eta_5)\s(\eta_6)]_+[\s(\eta_7)\s(\eta_8)]_+\ra \nn
|111\ra &\equiv& \la [\s(\eta_1)\s(\eta_2)]_-[\s(\eta_3) \s(\eta_4)]_-[\s(\eta_5)\s(\eta_6)]_-[\s(\eta_7)\s(\eta_8)]_-\ra  ,
\eeqa
where the subscript $\pm$ of the pair $[\s(\eta_i) \s(\eta_{i+1})]_\pm$ denotes the fermion parity of the fusion channel and is $+$ if the state
of the pair is $|0\ra$ and  $-$ if the state is $|1\ra$.
If we now transport adiabatically the Ising anyon with coordinate $\eta_7$ along a complete loop around that with coordinate 
$\eta_6$ this is equivalent to two subsequent applications of two  braid generators $B^{(8,+)}_6$ (both in the counter-clockwise 
direction) over the multiplet defining the basis (\ref{3qubits}).
Taking the explicit expression for the matrix $B^{(8,+)}_6$, where we have chosen for 
concreteness the  positive parity representation \cite{ultimate}  from Ref.~\cite{ultimate}, we obtain the following 
 monodromy matrix acting over the 8-fold degenerate multiplet (\ref{3qubits})
\beq \label{X_3}
\left(B^{(8,+)}_6\right)^2 = \left[ 
\matrix{0 & 1 & 0  & 0 & 0 & 0 & 0 & 0 \cr 
1 & 0 & 0  & 0 & 0 & 0 & 0 & 0 \cr
0 & 0 & 0  & 1 & 0 & 0 & 0 & 0 \cr
0 & 0 & 1  & 0 & 0 & 0 & 0 & 0 \cr
0 & 0 & 0  & 0 & 0 & 1 & 0 & 0 \cr
0 & 0 & 0  & 0 & 1 & 0 & 0 & 0 \cr
0 & 0 & 0  & 0 & 0 & 0 & 0 & 1 \cr
0 & 0 & 0  & 0 & 0 & 0 & 1 & 0}
\right] .
\eeq
This result illustrates why non-Abelian statistics is so interesting--by simply  exchanging two Ising anyons we
obtained a statistical `phase' which is not diagonal--it is a non-trivial non-diagonal statistical matrix. Other exchanges generate other
non-diagonal matrices acting on the same multiplet (\ref{3qubits}) and in general these non-diagonal statistical matrices do not 
commute, hence the name non-Abelian statistics.

The non-Abelian statistics  is definitely a new fundamental concept in two-dimensional particle physics,  which is interesting on its own.
However, it might also have a promising application in quantum computation. Notice that in quantum information language the matrix
(\ref{X_3}) is an implementation of the quantum NOT gate $X$ on the third qubit \cite{nielsen-chuang}, i.e., 
\[
\left(B^{(8,+)}_6\right)^2= X_3 = \Id_2\otimes\Id_2 \otimes X .
\] 
 Because information is encoded globally in degenerate multiplets of quantum states with $n$-anyons at fixed positions
  no local interaction can change or corrupt this information.
 In other words, quantum information is hidden from its enemies (noise and decoherence) which are due to local interactions, 
however, it is hidden even from us. In order to read this information non-local topologically nontrivial operations are needed 
\cite{sarma-freedman-nayak,sarma-RMP}.
This leads to the so-called  \textit{topological protection} of the encoded information and its processing. 
For example, for Ising anyons, the unprecedented precision of quantum information processing is due to the 
exponentially small probability for accidental creation of quasiparticle-quasihole pairs which is expressed in terms of the temperature 
and the experimentally estimated energy gap $\Delta\approx 500$ mK resulting in 
\[
 \mathrm{Error \ rate} \simeq  \left(\frac{k_B T}{\Delta}\right)
\exp\left(-\frac{\Delta}{k_B T}\right) < 10^{-30}
\]
for temperatures below 5 mK \cite{sarma-freedman-nayak}.

Given that the new concept of non-Abelian statistics is so interesting a natural question arises how could it be discovered.
Experiments with fractional quantum Hall (FQH) states, such as the $\nu_H=5/2$ state in which non-Abelian anyons are expected to exist,
 are conducted in extreme conditions and are difficult and expensive.
On top of that it appeared that there are other candidate FQH states, such as the 331 state, 
 having the same electric properties and identical patterns of Coulomb blockaded 
conductance peaks  but without non-Abelian anyons\cite{nayak-doppel-CB}. 
Therefore the conductance spectrometry \cite{stern-CB-RR,CB,thermal} of 
single-electron transistors, which was expected to detect non-Abelian statistics experimentally, is not sufficient to do that.

A possible resolution of this problem could be to measure the  thermoelectric characteristics of the Coulomb-blockaded 
FQH islands. The thermoelectric conductance, thermopower and especially the thermoelectric power factor 
\cite{viola-stern,NPB2015,NPB2015-2} considered here
could be the appropriate tools for detecting non-Abelian anyons, should they exist in Nature.
\section{The Pfaffian quantum Hall state and TQC with Ising anyons }
The Pfaffian FQH state, also known as the Moore--Read state \cite{mr} is the most promising candidate to describe the  
Hall state with filling factor $\nu_H=5/2$ observed experimentally in the second Landau level \cite{sarma-RMP}. 
Because the quasiparticle excitations in the Pfaffian state are known to be  non-Abelian \cite{mr} the 
 Hall state with  $\nu_H=5/2$ appears to be the most stable FQH state (with highest energy gap) in which  
non-Abelian anyons could be detected.

This  quantum Hall state is routinely observed in ultrahigh-mobility samples \cite{eisen,pan-xia-08,choi-west-08}
and is believed to be in the universality class of the Moore--Read state  \cite{mr} whose CFT is $\widehat{u(1)}\times \mathrm{Ising}$.  
The peculiar topological properties of non-Abelian anyons give some hope that non-Abelian statistics might be easier to be 
observed than the Abelian one \cite{stern-halperin}.
\subsection{TQC scheme with Ising anyons: single qubit construction }
\begin{figure}[htb]
\centering
\includegraphics*[bb=25 560 562 820,width=\textwidth]{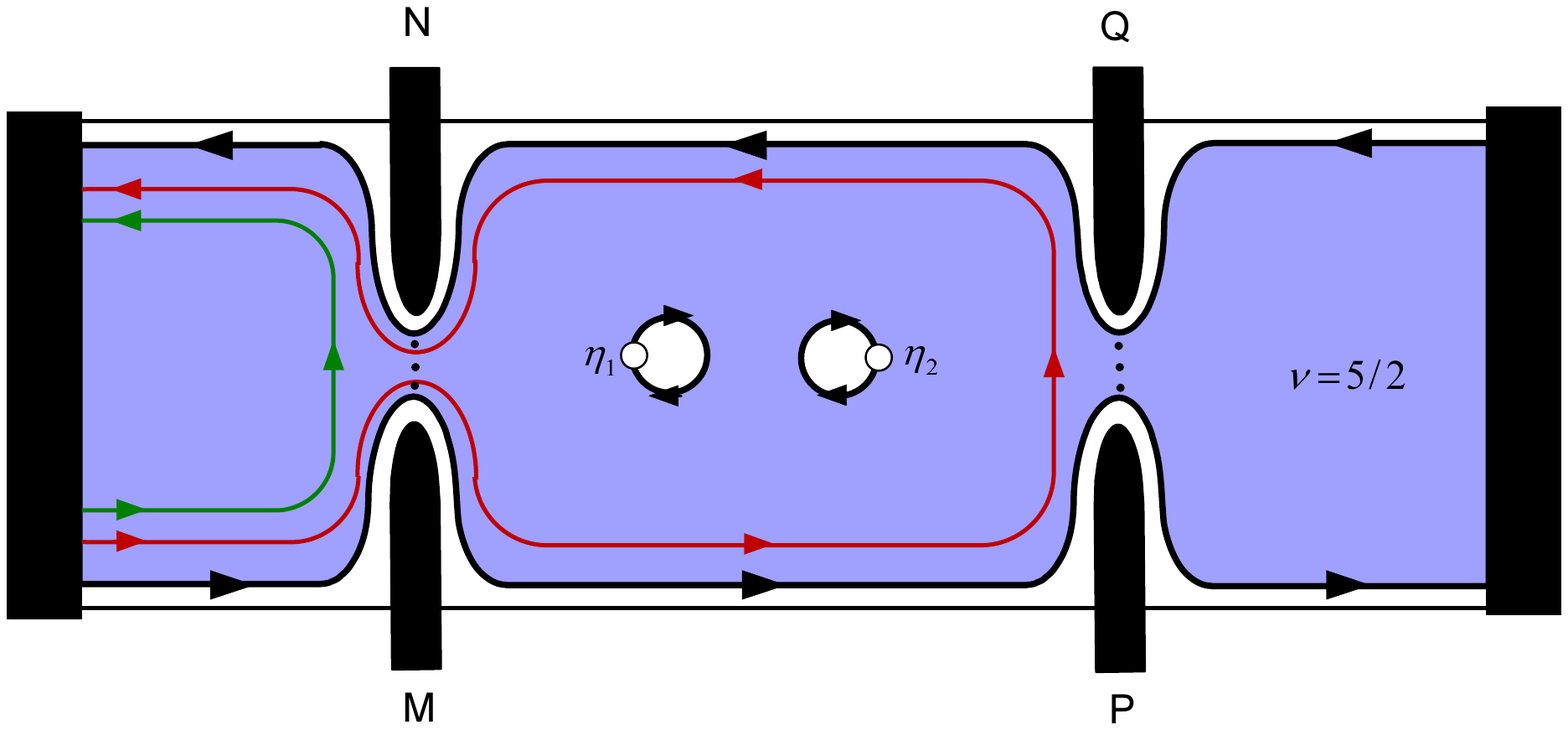}  
\caption{The qubit scheme of Das Sarma et. al. Only two antidots are shown and there are two anyons with 
coordinates $\eta_1$ and $\eta_2$ localized on the antidots \label{fig:qubit}}
 \end{figure}

The TQC scheme proposed by Das Sarma et al. \cite{sarma-freedman-nayak}
is based on a $\nu_H=5/2$ FQHE sample with 4  non-Abelian quasiparticles at fixed positions $\eta_a$ localized on 4 quantum antidots
as shown in Fig.~\ref{fig:qubit}.
The $N$-electron wave function describing the  sample with 4 anyons with coordinates $\eta_1, \ldots, \eta_4$ can be expressed 
as a  CFT correlation function \cite{mr}
\[
 \Psi_{\mathrm{qubit}} (\eta_1,\eta_2, \eta_3,\eta_4,\{ z_i \} )=
 \la \s(\eta_1)\s(\eta_2)\s(\eta_3)\s(\eta_4)  \prod_{i=1}^N  \psi_{\el}(z_i) \ra,
\]
 where $\psi_{\el}(z)$ is the field operator of the physical electron at position $z$ in the coordinate plane and 
 $\s(\eta)$ is the  non-Abelian Ising anyon at position $\eta$. Following Eq.~(\ref{encoding}) the qubit is in the state 
 $|0\ra$ if the two anyons  $\s(\eta_1)$ and $\s(\eta_1)$  fuse together to the unit operator $\Id$ while
 it is   in the state  $|1\ra$ if they fuse together to the Majorana fermion  $\psi$ and the state could be measured 
 by interferometric measurement of the conductance. For more details on the TQC scheme of Das Sarma et al. see 
Refs.~\cite{sarma-freedman-nayak, sarma-RMP,TQC-NPB}.
 
The quantum gates in the TQC scheme with Ising anyons are implemented by braiding, i.e., by adiabatic exchange of the 
antidots on which the Ising anyons are localized. The braiding of the 4 anyons generate two finite two-dimensional representations 
 of the braid group $\B_4$,  which are characterized by the positive or negative total fermion parity of the 4 anyon fields $\s$ \cite{ultimate}.
Denoting the braid matrices in the positive-parity representations as 
$B^{(4,+)}_1=R_{12}^{(4)}$, $B^{(4,+)}_2=R_{23}^{(4)}$ and $B^{(4,+)}_3=R_{34}^{(4)}$, 
where e.g. $R_{23}^{(4)}$ denotes the matrix representing the exchange of the anyons with coordinates 
$\eta_2$ and $\eta_3$,  we can write them explicitly as  \cite{ultimate}
\beq \label{B4+}
B_{1}^{(4,+)}=\left[ \matrix{1 & 0 \cr 0 & i}\right],
\quad
B_{2}^{(4,+)}=
\frac{\e^{i\frac{\pi}{4}} }{\sqrt{2}}
\left[ \matrix{\ \    1 & -i \cr -i & \ \ 1}\right], \quad
B_{3}^{(4,+)}=\left[ \matrix{1 & 0 \cr 0 & i}\right]
\eeq
and similarly for the negative-parity representation (with the corresponding notation)
\beq \label{B4-}
B_{1}^{(4,-)}=\left[ \matrix{1 & 0 \cr 0 & i}\right],
\quad
B_{2}^{(4,-)}=
\frac{\e^{i\frac{\pi}{4}} }{\sqrt{2}}
\left[ \matrix{\ \    1 & -i \cr -i & \ \ 1}\right], \quad
B_{3}^{(4,-)} = \left[ \matrix{i & 0 \cr 0 & 1}\right] .
\eeq
Multi-qubit states are realized by adding more pairs of Ising anyons.  Since each qubit state is encoded into one pair of anyons 
$(\s \s)_\pm$, where the subscript $\pm$ denotes the total fermion parity of the pair,
\begin{svgraybox}
 an $n$-qubit state can be realized by $2n+2$ Ising anyons $\s$ localized on $2n+2$ antidots.  Therefore, the exchanges of the $2n+2$ Ising anyons in the $n$-qubit register generate
representations of the braid group $\B_{2n+2}$ which are again characterized by the total fermion parity of the $\s$ fields.
\end{svgraybox}
The last 2 anyons (or any other chosen pair) among the $2n+2$ anyons are inert because they carry no information--their only 
purpose is to compensate the total fermion parity so that the CFT correlation function is non-zero.

Interestingly enough, it was found in Ref.~\cite{ultimate}, that  the generators $B_j^{(2n+2,\pm)}$, $j=0,\ldots, 2n+1$, of the 
  braid group $\B_{2n+2}$ can be expressed in terms of the generators $B_j^{(2n,\pm)}$, $j=0,\ldots, 2n-1$, 
for $\B_{2n}$ due to the following \textit{recursive relations} 
 \beqa \label{recursive}
B_j^{(2n+2,+)} &=& B_j^{(2n+2,-)} \quad \quad \qquad \textrm{for} \quad  1\leq j \leq 2n \nn
B_j^{(2n+2,\pm)} &=& B_j^{(2n,\pm)} \otimes \Id_2 \qquad  \quad \textrm{for} \quad 1\leq j \leq 2n-3 \nn
B_j^{(2n+2,\pm)} &=& B_{j-2}^{(2n,\pm)} \oplus B_{j-2}^{(2n,\mp)} \quad \textrm{for} \quad 3\leq j \leq 2n+1  .
\eeqa
Using the recursion relations (\ref{recursive}) together with Eqs.~(\ref{B4+}) and (\ref{B4-}) we can find explicitly all braid generators.
This will allow us to build almost all quantum gates as products of the braid generators and implement them by subsequent 
braiding of Ising anyons. We emphasize here that all quantum gates which can be implemented by  braiding of non-Abelian anyons
are topologically protected hardware for topological quantum computers \cite{sarma-RMP}.
\subsection{Single-qubit gates: The Pauli $X$ gate}
\begin{figure}[htb]
\centering
\includegraphics*[bb=0 340 595 520,width=9cm]{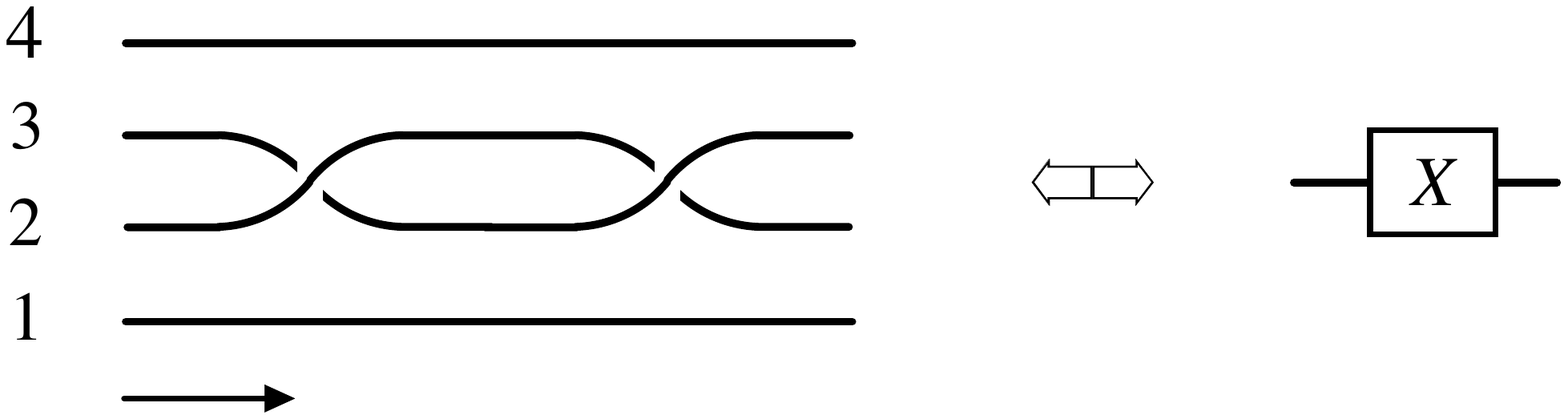} 
\caption{Braiding diagram for the Pauli $X$ gate and its quantum computation symbol. \label{fig:X} }
\end{figure}

The first gate which has been implemented by braiding of Ising anyons \cite{sarma-freedman-nayak} is the NOT gate \cite{nielsen-chuang}
 which is usually denoted as the Pauli $X$ matrix
\[
X \equiv  (R_{23})^2= \left(B_{2}^{(4,\pm)} \right)^2 = \left[ \matrix{0 & 1 \cr 1 & 0}\right].
\]
It can be realized by taking the anyon with coordinate $\eta_2$ along a complete loop around the anyon with coordinate $\eta_3$.
Using Eqs.~(\ref{B4+}) and (\ref{B4-}) we can easily check  that the square of the generator $B_{2}^{(4,\pm)}$ indeed implements 
the NOT gate and this process corresponds to the braid diagram given in Fig.~\ref{fig:X}.
\subsection{The Hadamard gate}
\label{sec:H}
\begin{figure}[htb]
\centering
\includegraphics*[bb=0 355 595 500,width=9cm]{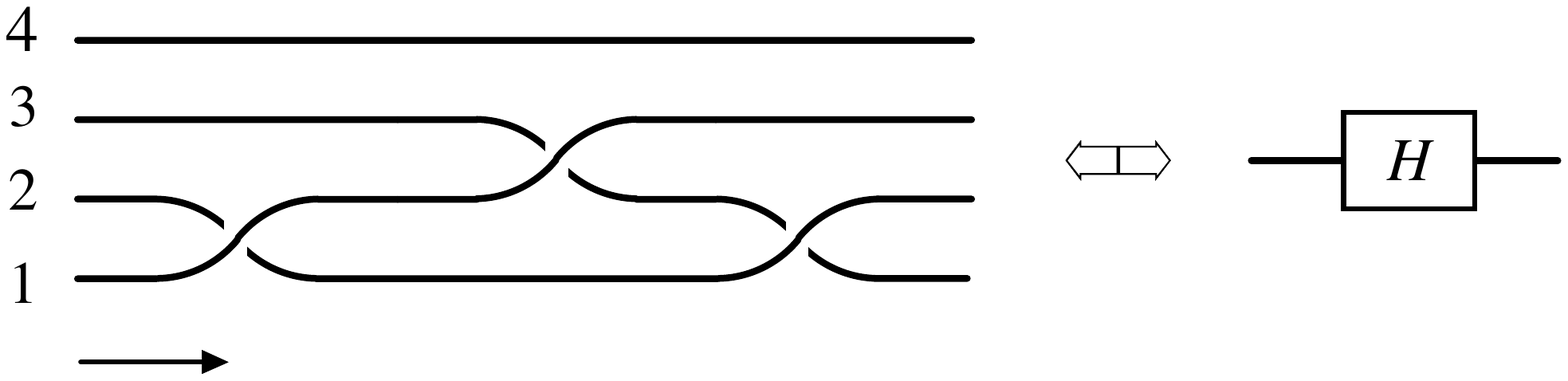} 
\caption{Hadamard gate implemented by braiding and its quantum computation symbol. \label{fig:H}}
\end{figure}

Another important single-qubit gate is the Hadamard gate  \cite{nielsen-chuang}. It can be implemented by 
braiding \cite{TQC-PRL,TQC-NPB}
as follows: first exchange the anyons with coordinates $\eta_1$ and $\eta_2$, then exchange the anyons with coordinates 
$\eta_2$ and $\eta_3$ and finally exchange again the anyons with coordinates $\eta_1$ and $\eta_2$, as shown in the braid diagram
in Fig.~\ref{fig:H}
which is equivalent to exchanging counter-clockwise  first the anyons with coordinates $\eta_1$ and $\eta_3$ and then 
taking the anyon with coordinate $\eta_2$  along a complete loop around that with coordinate $\eta_1$, i.e. 
\beq \label{H}
H\simeq \left(R_{12}\right)^2 R_{13} = R_{12} R_{23} R_{12}  = B_{1}^{(4,\pm)} B_{2}^{(4,\pm)}B_{1}^{(4,\pm)}=
\frac{\e^{i\frac{\pi}{4}} }{\sqrt{2}}
\left[ \matrix{1 & \ \ \  1 \cr 1 & -1}\right],
\eeq
where $R_{i j}$ in the notation of Refs.~\cite{TQC-PRL,TQC-NPB} is the operation representing the counter-clockwise 
exchange of anyons with coordinates $\eta_i$ and $\eta_j$.
\subsection{The phase  gate $S$ }
We emphasize here again that in order to build a universal topological quantum computer we need to be able to implement by braiding
the three universal gates given in Eq.~(\ref{discrete}). The Hadamard gate $H$ has been implemented by braiding of Ising anyons in 
Sect.~\ref{sec:H}. The only single-qubit that need to be implemented by braiding is the $T$ gate. Unfortunately, the $T$ gate cannot be 
implemented by braiding of Ising anyons \cite{clifford}, which means that the Ising TQC is not universal. However, the $T$ gate might 
be implemented in another way which is not topologically protected by exponential suppression of pair activation due to the energy gap
but still with a low error rate. On the other hand, braiding Fibonacci anyons can be used to build a truly universal topological quantum 
computer \cite{sarma-RMP}.

\begin{figure}[htb]
\centering
\includegraphics*[bb=0 330 595 530,width=9cm]{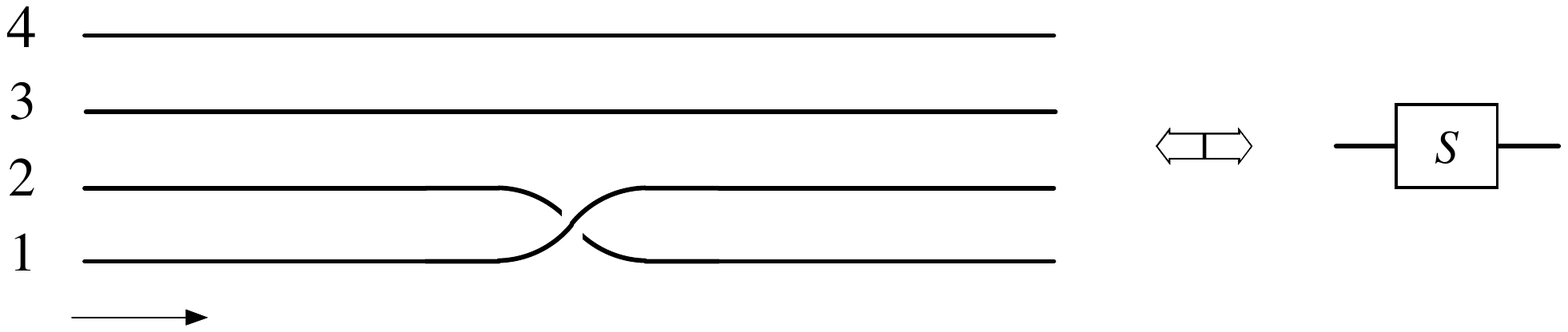}
\caption{The phase gate $S$ realized by braiding of Ising anyons and its quantum computation symbol. \label{fig:S}}
\end{figure}

Nevertheless, the Ising TQC is important  enough because of its stability so it is interesting to know what is the maximal subgroup
of $SU(2^n)$ of quantum gates which can be implemented by braiding Ising anyons. The answer is that this is a subgroup of the Clifford group
\cite{clifford}, the group that preserves the Pauli group \cite{clifford} and therefore plays a central role in the error-correcting algorithms. 
It is worth mentioning that all quantum gates which can be implemented by braiding of Ising anyons are Clifford gates.
However, not all Clifford bates are realizable by braiding. Although the $T$ gate is not implementable by braiding its square $S=T^2$ is.
This important Clifford gate \cite{nielsen-chuang} could indeed be realized by braiding of Ising anyons as follows \cite{TQC-NPB}
\[
S=T^2 \equiv R_{12}= R_{34} =B_{1}^{(4,\pm)}= \left[ \matrix{1 & 0 \cr 0 & i}\right].
\]
and the corresponding braid diagram is shown in Fig.~\ref{fig:S}.
The fact that all Clifford gates could be realized by braiding of Ising anyons is promising because of the 
development of the error-correcting codes \cite{nielsen-chuang} and might eventually compensate the lack of complete
topological protection of the quantum gates constructed in the Ising topological quantum computer.
\subsection{Two-qubits construction}
\begin{figure}[htb]
\centering
\includegraphics*[bb=80 570 510 700,width=11cm]{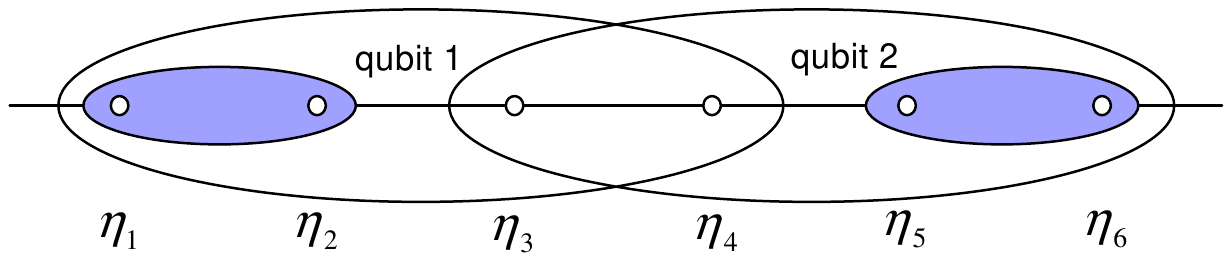}
\caption{Two qubits constructed from 6 Ising anyons $\s(\eta_i)$ at fixed positions $\eta_i$. \label{fig:2qubits}}
\end{figure}

As mentioned earlier the $n$-qubit states belong to the tensor product of $n$ single-qubit spaces, each of which is two dimensional 
with the condition that the $n$-qubit states must be normalized. Therefore, the dimension of the $n$-qubit Hilbert spaces is $2^n$.
On the other hand, it is known that the space of the correlation functions of the Ising model with $2n$ Ising anyons $\s$ has dimension
dim $\H_{2n} =2^{n-1}$, so that $2n+2$ Ising anyons could be used to represent a general $n$-qubit state. Taking into account the 
quantum information encoding into the Ising anyon pairs fusion channel,  specified in Eq.~(\ref{encoding}), we consider the following
6 Ising anyons realization of the two-qubit computational basis in the Ising TQC
\beqa
&&|00\ra \equiv \la \s_+ \s_+\s_+\s_+\s_+\s_+\ra, \quad
\! |01\ra \equiv \la \s_+ \s_+\s_+\s_-\s_+\s_-\ra \nn
&&|10\ra \equiv \la \s_+ \s_-\s_+\s_-\s_+\s_+\ra, \quad
|11\ra \equiv \la \s_+ \s_-\s_+\s_+\s_+\s_-\ra  , \nonumber
\eeqa
as shown in Fig.~\ref{fig:2qubits}, where the first pair of Ising anyons (with coordinates $\eta_1$ and $\eta_2$) correspond to the first qubit, the third pair of Ising anyons 
 (with coordinates $\eta_5$ and $\eta_6$) represents the second qubit
and the two Ising anyons between them  (with coordinates $\eta_3$ and $\eta_4$) is an inert pair which, on one side compensates 
total fermion parity so that the correlation function is non-zero, while on the other side creates topological entanglement
between the two Ising qubits which can be used to construct by braiding some entangling two-qubit gates, such as the CNOT gate.
\subsection{Two-qubit gates: the Controlled-NOT gate }
The two-qubit CNOT gate is one of the most important resource for quantum computation \cite{nielsen-chuang} because this
entangling gate can transfer information from one qubit to another. Moreover, due to the CNOT gate, one can express any $n$-qubit gate,
which can be written as a unitary matrix from the group $SU(2^n)$, as a product of single-qubit 
gates (two-dimensional unitary matrices in tensor product with $n-1$ unit matrices $\Id_2$  completing the dimension of the matrix to 
$2^n$) with
two-qubit gates  (four-dimensional unitary matrices in tensor product with $n-2$ unit matrices $\Id_2$  unit matrices completing 
the dimension of the matrix to $2^n$), see Ref.~\cite{nielsen-chuang}.
In this subsection we will demonstrate that the CNOT gate can be constructed explicitly in terms of 6-anyon elementary braidings using 
the notations of Ref.~\cite{TQC-NPB} in which $R_{i, i+1} = B_{i}^{(6,+)}$ are the 
representations of the generators  of the braid group $\B_6$ corresponding to the elementary braids of 6 Ising anyons.

\begin{figure}[htb]
\centering
\includegraphics*[bb=10 340 580 490,width=\textwidth]{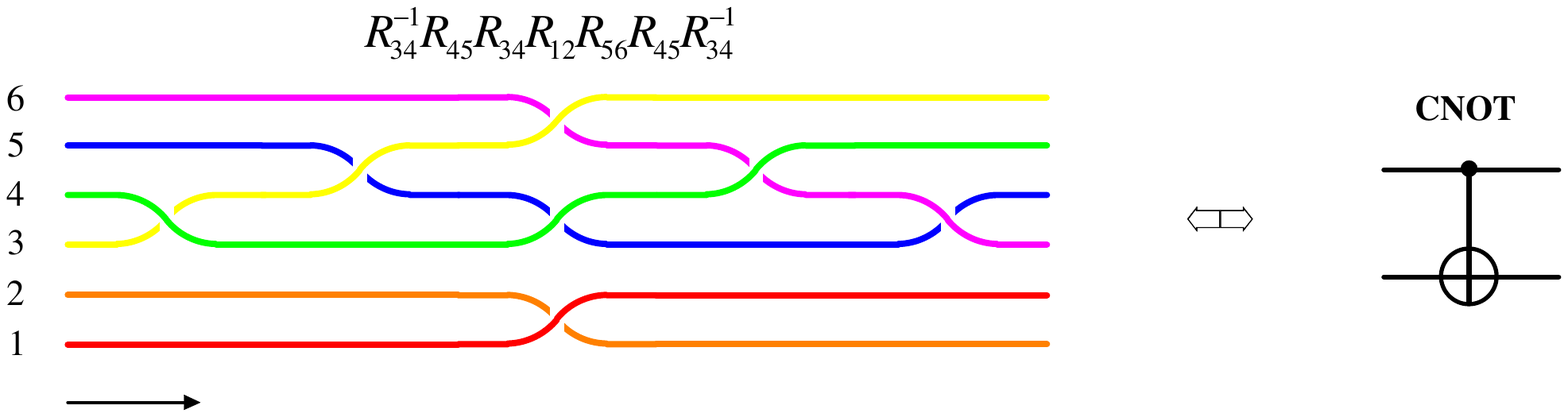}  
\caption{The CNOT gate realized by braiding of 6 Ising anyons and its quantum computation symbol. \label{fig:CNOT}}
\end{figure}

Our strategy \cite{TQC-PRL,TQC-NPB} is  to use first the well known connection \cite{nielsen-chuang} between the CNOT and the 
Controlled-Z  (CZ)  gate, which is the diagonal matrix with the following elements on the diagonal CZ$ \ =\mathrm{diag}(1,1,1,-1)$,
in terms of the Hadamard gate\footnote{recall that the single-qubit Hadamard gate has been constructed in Eq.~(\ref{H})} 
acting on the second qubit $H_2=\Id_2\otimes H$ and then to try to construct the diagonal gate CZ 
in terms of the diagonal braid generators. It is not difficult to check \cite{TQC-NPB} 
that $R_{12} R_{34}^{-1} R_{56} =\mathrm{diag}(1,1,1,-1)$ and  $R_{56} R_{45} R_{56} =H_2$ so that one realization of CNOT is 
\beqa
\mathrm{CNOT} =
H_2 \  \mathrm{CZ} \ H_2
 =  R_{56} R_{45} R_{56}^{-1} R_{34}^{-1} R_{12}R_{45} R_{56}    
 \simeq
 \left[
\matrix{1 & 0 & 0 &  0 \cr 0 & 1 & 0  &  0 \cr 0 & 0 & 0 & 1
	\cr 0 & 0 & 1 &  0} \right]  . \nonumber
\eeqa
Alternatively the CNOT can be constructed by another combination of 6-anyon braids, namely
\[
\mathrm{CNOT}=R_{34}^{-1} R_{45} R_{34}R_{12}R_{56} R_{45}  R_{34}^{-1} ,
\]
which is graphically shown on Fig.~\ref{fig:CNOT}
\begin{svgraybox}
What is remarkable in this construction of the CNOT gate by braiding 6 Ising anyons is that it is completely topologically protected
and consist only 7 elementary braids. Unfortunately, the construction of the embedding of the two-qubit CNOT into 
Ising systems with more than 2 qubits is not possible by braiding Ising anyons only \cite{clifford} which is a clear limitation 
of the Ising TQC.
\end{svgraybox}
Despite this limitation of the Ising TQC it is still worth investigating it since all gates that can be implemented in a fully
topologically protected way by braiding Ising anyons are Clifford gates which is very important for the quantum error correcting 
codes \cite{nielsen-chuang}.

In the rest of this paper we will focus on how non-Abelian Ising anyons could be detected experimentally if they exist.
We start by the description of the Coulomb blockaded quantum Hall islands which are equivalent to
 single-electron transistors and review their conductance spectroscopy and then we will consider
the thermoelectric characteristics of these islands.
\section{Coulomb-blockaded quantum Hall islands: QD and SET}
\begin{figure}[htb]
\centering
\includegraphics[bb=20 280 575 550,clip,width=\textwidth]{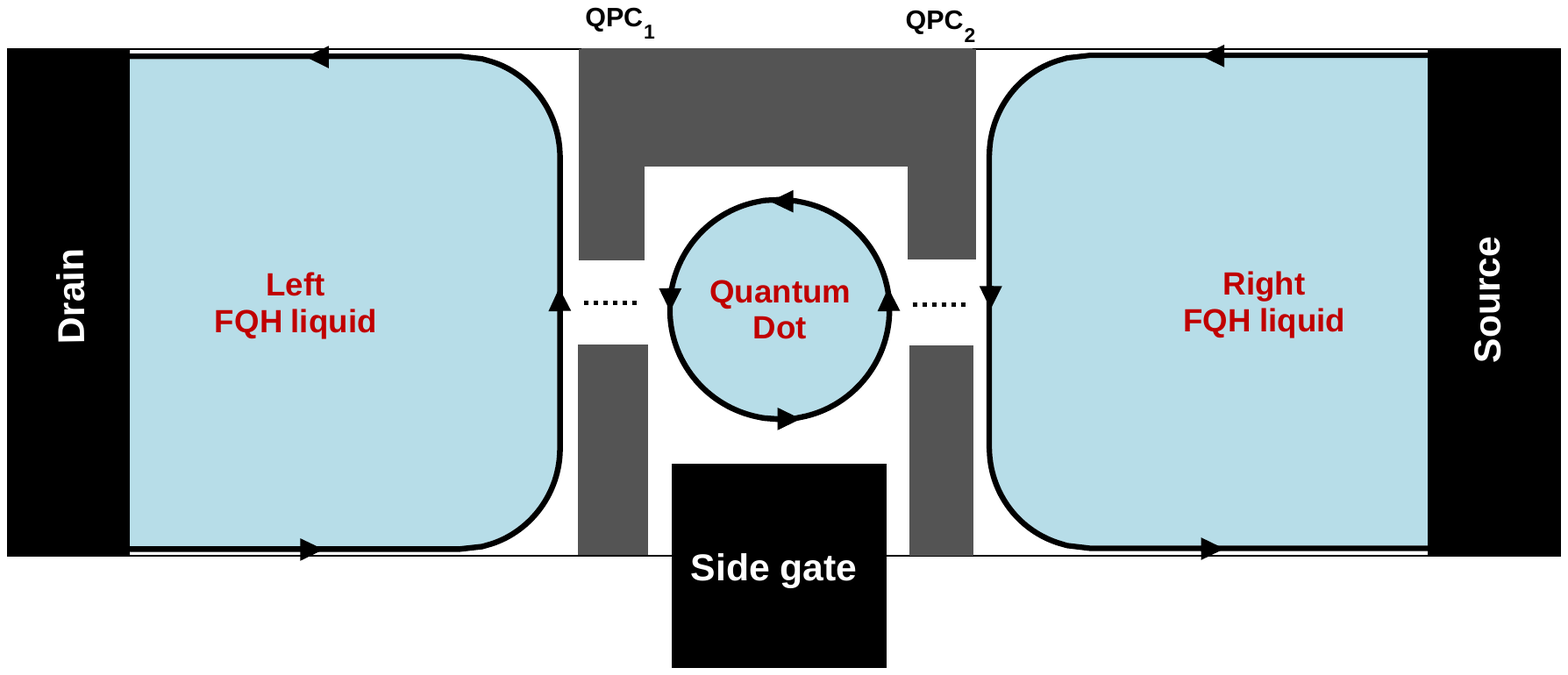}
\caption{Single-electron transistor. \label{fig:SET}}
\end{figure}

The experiments, which are expected to shed more light on the nature of the quasiparticle excitations in FQH states,
are performed in single-electron transistors (SET) constructed as Coulomb blockaded islands, or quantum dots, equipped 
with drain, source and side gates as shown in Fig.~\ref{fig:SET}. 
The quantum dot (QD) is realized by splitting a larger FQH bar with the help of two quantum point contacts (QPC) and it is
assumed that the left- and right- FQH liquids are much larger in size than the QD, so that the energy spacing in the 
left- and right- FQH liquids is much smaller \cite{matveev-LNP} than the energy spacing in the QD, which is proportional to
$\hbar 2\pi v_F/L$, where $v_F$ is the Fermi velocity of the electrons on the QD's edge and $L$ is the circumference of the 
edge. When $L$ is small enough the energy spacing is big and the QD's energy levels are discrete. In this setup, for low 
temperatures, 
electron can tunnel from the left FQH liquid to the QD and then to the right FQH liquids only when the chemical potentials 
of the left and right FQH liquids are aligned with some of the discrete energy levels of the QD. On the other hand, changing the 
potential of the side gate, which is capacitively coupled to the QD, continuously shifts up or down the energy levels of the QD.
Therefore, measuring the electric conductance as a function of the gate voltage can be viewed as precise energy 
level spectroscopy  of the QD and this can be used to distinguish FQH systems.
\subsection{Coulomb island's conductance--CFT approach}
Although the electric conductance is a non-equilibrium quantity in the regime when the tunneling through the SET 
 is weak we can use the linear response approximation and express the conductance as an equilibrium thermal average.
For this purpose we are going to use the Grand canonical partition function for a FQH disk, which represents a QD
in which the bulk is inert, i.e., the electric charge carriers in the bulk are localized and do not contribute to the electric current, 
while the edge is mobile in the sense that charge carriers are mobile. The edge of such strongly correlated electron systems 
in the FQH regime could be described  by an effective rational unitary CFT \cite{fro-stu-thi,cz,CFT-book}. 
The Grand partition function for the edge of the FQH disk can be written as
\beq \label{Z}
Z(\t,\z) = \mathrm{tr}_{ \H_{\mathrm{edge}}} \ \e^{-\beta (H-\mu N)} 
= \mathrm{tr}_{ \H_{\mathrm{edge}}} \ \e^{2\pi i \t (L_0 -c/24)} e^{2\pi i \z J_0},
\eeq
 where the Hamiltonian $H=\hbar\frac{2\pi v_F}{L} \left(L_0-\frac{c}{24}\right)$ is expressed in terms of the zero mode $L_0$ of the 
 Virasoro stress-energy tensor \cite{CFT-book} with central charge $c$, the electron number operator  $N=-\sqrt{\nu_H} J_0$  
 is expressed in terms of the zero mode of the $\uu$ current and $\nu_H$ is the (quantum) Hall filling factor.
The trace is taken over the Hilbert space    $\H_{\mathrm{edge}}$ for the  edge states and might depend on the type and number of the 
quasiparticles which are localized in the bulk. The modular parameters   $\z$ and $\t$ of the rational CFT  \cite{CFT-book}, which 
appear in Eq.~(\ref{Z}) are related to the temperature $T$ and chemical potential $\mu$ as follows
\beq \label{modular}
\t=i\pi\frac{T_0}{T}, \quad 
T_0=\frac{\hbar v_F }{\pi k_B L}, \quad \z= i\frac{\mu}{2\pi  k_B T} =\frac{\mu}{\Delta\epsilon} \t , \quad \Delta\epsilon=\hbar\frac{2\pi v_F}{L},
\eeq
where $v_F$ is the Fermi velocity on the edge,  $L$ is the edge's circumference and $k_B$ is the Boltzmann constant.

When the FQH disk is threaded by perpendicular magnetic field $B$ the edge is affected by the flux of this field 
$B A$, where $A$ is the area of the disk and the concrete type of the field $B$ is not important. Therefore we can assume that
the disk is threaded by $B_{\mathrm{total}}=B_0+B$, where $B_0$ is a constant homogeneous magnetic field, corresponding to the 
center of a FQH plateau, while $B$ is magnetic field of the Aharonov--Bohm (AB) type. This is very convenient because  the
AB flux can be treated analytically \cite{NPB-PF_k} and  the disk CFT partition function $ Z_{\phi}(\t,\z ) $ in presence of 
(dimensionless) AB flux $\phi$ is simply obtained by a shift in the modular parameter $\z \to \z +\phi \t$ \cite{NPB-PF_k}, i.e.
 \beq \label{shift}
 Z_{\phi}(\t,\z ) = Z(\t,\z +\phi\t), \quad  \phi= \frac{e}{h}(B_{\mathrm{total}}-B_0)A ,
\eeq
where $h$ is the Plank constant. Interestingly enough, the variation of the side-gate voltage is affecting the QD in the same 
way \cite{NPB2015} as  the AB flux $\phi$ through the \textit{externally induced electric charge}  $Q_{\mathrm{ext}}$ on QD,  
which changes continuously  with the gate voltage $V_g$
\beq \label{V_g}
-\frac{C_g V_g}{e} \equiv \nu_H \phi =Q_{\mathrm{ext}} ,
\eeq
where $C_g$ is the capacitance of the gate. Therefore we can use the partition function (\ref{shift}) to compute various
 thermodynamic quantities as functions of the gate voltage $V_g$.

 The Grand potential on the edge, in presence of AB flux $\phi$ or gate voltage $V_g$ defined in Eq.~(\ref{V_g}), can be
 expressed as 
\beq \label{omega_phi}
\Omega_{\phi}(T,\mu)=-k_B T \ln Z_{\phi}(\t,\z),
\eeq
where $Z_\phi$ is defined in Eq.~(\ref{shift}) and the thermal average of the electron number can be 
computed by  \cite{thermal}
\beqa \label{N}
\la N_\el (\phi)\ra_{\beta,\mu_N} &=& {-\frac{\partial \Omega_\phi(\beta,\mu_N)}{\partial \phi} } + {\nu_H\phi}
 +\nu_H\left(\frac{\mu_N}{\Delta \epsilon} \right) \nn
 &=&\nu_H\left(\phi+ \frac{\mu_N}{\Delta \epsilon} \right) +\frac{1}{2\pi^2} \left(\frac{T}{T_0}\right) \frac{\partial }{\partial \phi}
 \ln Z_\phi(T,\mu_N) ,
\eeqa 
 where $\mu_N$ is the chemical potential of a QD with $N$ electrons. Similarly, 
 the edge conductance $G_{\mathrm{is}}$ of the Coulomb blockade island,  in presence of AB flux $\phi$ or gate voltage 
$V_g$ defined in Eq.~(\ref{V_g}), can be computed by \cite{thermal}
\beq \label{G}
G_{\mathrm{is}} (\phi)=\frac{e^2}{h}
\left( \nu_H +\frac{1}{2\pi^2} \left(\frac{T}{T_0} \right)\frac{\partial^2 }{\partial \phi^2}  \ln Z_{\phi}(T,0)\right) .
\eeq
As an example the profile of the conductance for the Pfaffian FQH disk is given in Sect.~\ref{sec:Pf}. 
\begin{svgraybox}
It is important to emphasize that the conductance peak patterns of SETs, when gate voltage is varied, are not sufficient to
distinguish different FQH states sharing the same $\uu$ part and therefore having the same electric properties
while differing in the neutral part of the CFT \cite{nayak-doppel-CB}. The way out is to compute some thermoelectric
characteristics of SETs, which are sensitive to the neutral sector of the CFTs and could eventually be used to
distinguish between different FQH states having the same electric properties.
\end{svgraybox}
\subsection{Thermopower: a finer spectroscopic tool  }
The thermopower, or the Seebeck coefficient \cite{matveev-LNP}, is defined as the  potential difference $V$ generated between the two 
leads of the SET when the temperatures $T_L$ and $T_R$ of the two leads is different and $\Delta T=T_R -T_L\ll T_L$, 
under the condition that $I=0$. Usually thermopower is expressed \cite{matveev-LNP} as the ratio $G_T/G$, where $G_T$ and $G$ are
the thermal and electric conductances respectively, however for a SET, both $G$ and $G_T$ are 0 in large intervals of
gate voltages, called the Coulomb valleys, so it is more convenient to use another expression for thermopower \cite{matveev-LNP}
\beq \label{S}
S \equiv  \left. -\lim_{\Delta T \to 0} \frac{V}{\Delta T}\right|_{I=0}=
-\frac{\la \varepsilon \ra}{eT}, \quad
\eeq
where $\la \varepsilon \ra$ is the average energy of the tunneling electrons through the SET. It is intuitively clear that the average tunneling 
energy can be expressed in terms of the total energies of the QD with $N+1$ electrons and of the QD with $N$ electrons as
\[
\la \varepsilon \ra^{\phi}_{\beta,\mu_N} = E^{\beta,\mu_{N+1}}_{\mathrm{QD}}(\phi)-  E^{\beta,\mu_{N}}_{\mathrm{QD}}(\phi)  ,
\]
where the total  QD energy (with $N$ electrons on the QD)  can be written (in the Grand canonical ensemble)  as
\[
E^{\beta,\mu_{N}}_{\mathrm{QD}}(\phi) =\sum_{i=1}^{N_0} E_i+\la H_{\mathrm{CFT}}(\phi)\ra_{\beta,\mu_{N}} .
\]
Here $E_i$, $i=1, \ldots , N_0$ are the occupied single-electron states in the bulk of the QD,  
and  $\la \cdots \ra_{\beta,\mu}$ is the Grand canonical average of $H_{\mathrm{CFT}}$ on the edge at inverse temperature
$\beta=(k_B T)^{-1}$ and chemical potential $\mu$.
 However, because we are working
with Grand canonical partition functions the difference of the thermal averages 
$\la N(\phi)\ra_{\beta,\mu_{N+1} } -  \la N(\phi)\ra_{\beta,\mu_{N} }$ of the electron numbers of the QDs with $N+1$ and $N$ electrons 
is not  1 for all values of $V_g$, as can be seen if we plot this difference using Eq.~(\ref{N}). Therefore, it is more appropriate  
to express the average tunneling energy as \cite{NPB2015} 
\beq \label{eps}
\la \varepsilon \ra^{\phi}_{\beta,\mu_N} = 
\frac{\la H_{\mathrm{CFT}}(\phi)\ra_{\beta,\mu_{N+1} }  -  \la H_{\mathrm{CFT}}(\phi)\ra_{\beta,\mu_{N} } }
{{\la N_{\mathrm{el}}(\phi)\ra_{\beta,\mu_{N+1} }  -  \la N_{\mathrm{el}}(\phi)\ra_{\beta,\mu_{N} } }}  .
 \eeq
All thermal averages in Eq.~(\ref{eps}) can be computed  within our CFT approach from the partition function (\ref{shift}) for the FQH 
edge in presence of AB flux $\phi$ or gate voltage $V_g$. For example, the electron number average can be computed
from Eq.~(\ref{N}), while the edge energy average can be computed from the standard Grand canonical ensemble expression
\beq \label{H2}
\la H_{\mathrm{CFT}}(\phi)\ra_{\beta,\mu_N} =\Omega_{\phi}(T,\mu_N)- T \frac{\partial \Omega_{\phi}(T,\mu_N)}{\partial T}  
 - \mu_N\frac{\partial  \Omega_{\phi}(T,\mu_N)}{\partial \mu}   
\eeq
 where $\Omega_{\phi}(T,\mu_N)$ is defined in Eq.~(\ref{omega_phi}) and the chemical potentials $\mu_N$ and  $\mu_{N+1}$
corresponding to QD with $N$ and $N+1$ electrons respectively are given by \cite{NPB2015}
\[
\mu_N=-\frac{1}{2}\Delta\epsilon, \quad \mu_{N+1}=\frac{1}{2}\Delta\epsilon,   
\]
where $\Delta\epsilon$ is defined in Eq.~(\ref{modular}).
One very interesting thermoelectric quantity is the \textit{thermoelectric power factor} $\P_T$, which is defined as the 
electric power $P$ generated by the temperature difference $\Delta T$ and can be expressed from Eq.~(\ref{S}) in terms 
of the thermopower $S$ and the electric conductance $G$ as \cite{NPB2015}
\beq \label{PF}
P=V^2 / R = {\P_T }(\Delta T)^2, \quad \P_T= S^2 G,
\eeq
where $R=1/G$ is the electric resistance of the CB island. In the next section we will calculate the power factor for SET in the 
Pfaffian FQH state.
\section{The Pfaffian state: comparison with the experiment at $\nu_H=2/3$  }
\label{sec:Pf}
A recent experiment \cite{gurman-2-3} brought a new hope for the possibility to measure some thermoelectric characteristics
of SETs and eventually detect ``upstream neutral modes'' in the fractional quantum Hall regime.
In this section we will calculate and plot the conductance, thermopower and  thermoelectric power factor for the Pfaffian
FQH state and will compare these quantities with the measured ones in the experiment \cite{gurman-2-3}.
It appears that there two distinct cases: even number of quasiparticles localized in the bulk and odd number,
which we shall consider separately.
\subsection{Thermopower for odd number of bulk quasiparticles} 
The disk partition function for the $\nu_H=5/2$ Pfaffian state with odd  number of quasiparticles localized  in the bulk can be 
written as \cite{NPB2015}
\beq\label{Z-odd}
Z^{\mathrm{odd}}(\tau,\zeta)=\left[K_1(\t,2\z;8)+K_{-3}(\t,2\z;8)\right]\ch_{1/16}(\t'),
\eeq
where $\t'=(v_n/v_c) \t$ is the modified modular parameter for the neutral partition functions
taking into account the anticipated difference in the  Fermi velocities $v_n$ and $v_c$ of the neutral and charged modes respectively.
The Luttinger liquid partition functions $K$ are defined (upto an unimportant $\z$-independent multiplicative factor) 
as \cite{NPB2015,CFT-book}
 \beq \label{K}
K_{l}(\t,\z; m) \propto \sum_{n=-\infty}^{\infty} q^{\frac{m}{2}\left(n+\frac{l}{m}\right)^2} 
\e^{2\pi i \z \left(n+\frac{l}{m}\right)},
\eeq
where  $q=\e^{2\pi i \t }= \e^{-\beta\Delta\epsilon}$, $\beta=(k_B T)^{-1}$. 
Because the partition function (\ref{Z-odd}) is a product of a $\z$-independent factor and a $\z$ dependent one the neutral 
partition function $\ch_{1/16}(\t')$ drops out of the calculations after taking the log in Eq.~(\ref{omega_phi}) and taking into account that 
differentiation  with respect to $\phi$ is the same as differentiation with respect to $\z$ due to Eq.~(\ref{shift}).
Thus the partition function for the $\nu_H=5/2$ Pfaffian state with odd  number of quasiparticles  in the bulk is
 the same as that for the Abelian $\nu_H=1/2$ Luttinger liquid ($R_c=1/2$) \cite{NPB2015}.
 
\begin{figure}[htb]
\centering
\includegraphics[bb=10 15 570 380,clip,width=10cm]{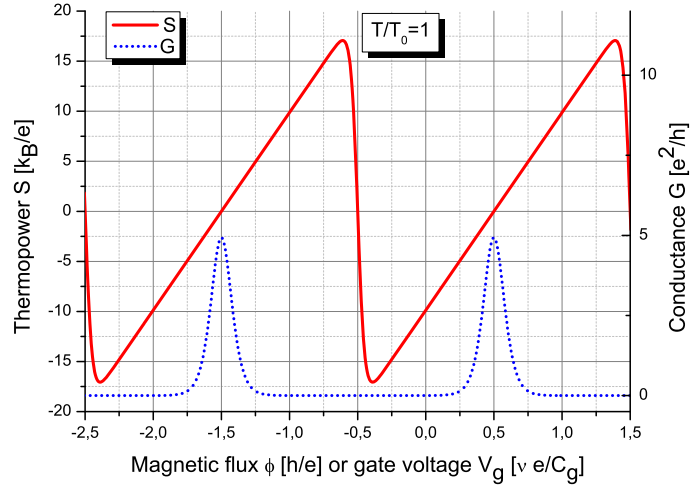}
\caption{Thermopower (left $Y$-scale) and conductance (right $Y$-scale)  for the $\nu_H=5/2$ Pfaffian state with 
odd  number of quasiparticles  in the bulk. 
\label{fig:TP}}
\end{figure}

 Now that we have the explicit partition function for the $\nu_H=5/2$ Pfaffian state with odd  number of quasiparticles  in the bulk
 we can compute the Grand canonical thermal averages  and using  Eqs.~(\ref{eps}), (\ref{N}),  (\ref{G}) and ~(\ref{H2})     
to compute the conductance and the thermopower which are plotted for $T=T_0$  in Fig.~\ref{fig:TP} as functions of the gate voltage.
It is interesting to note that conductance is zero in large intervals of gate voltage, which is called the Coulomb blockade, 
except at specific values of the gate voltage at which we observe conductance peaks.
The lower the temperature the sharper and narrower the conductance peaks. The thermopower on the other hand has a sawtooth form
whose jumps become vertical for $T \to 0$. The centers of the conductance peaks correspond to the zeros of the thermopower
like in metallic islands \cite{matveev-LNP}.

\begin{figure}[htb]
\centering
\includegraphics[bb=10 15 570 390,clip,width=10cm]{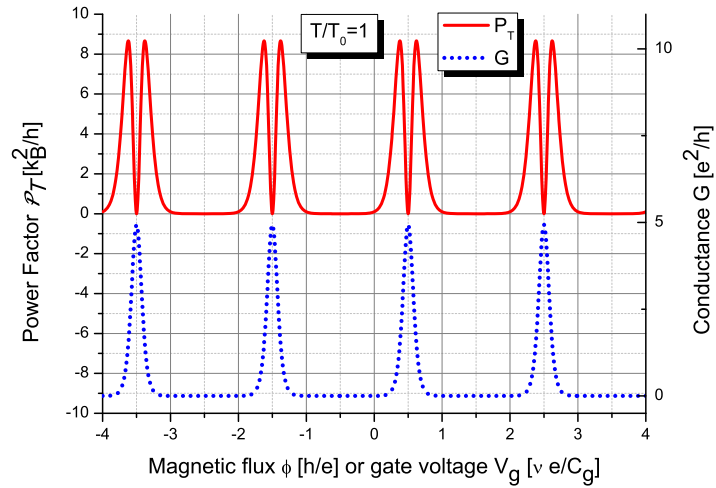}
\caption{Thermoelectric power factor (left) and conductance (right) of the Pfaffian state with odd number of bulk 
quasiparticles. \label{fig:PF}}
\end{figure}

Similarly, using the computed thermopower and Eq.~(\ref{PF}) we can calculate the thermoelectric power factor
for the (Moore--Read) Pfaffian state with odd number of bulk quasiparticles, which is plotted in Fig.~\ref{fig:PF}.
Because the thermopower vanishes at the centers of the conductance peaks the power factor shows very sharp
dips at these values of the gate voltage. 

It is very instructive to compare the theoretical calculation of the conductance, the thermopower and the power factor
for the Pfaffian state with odd number of bulk quasiparticles to the experiment \cite{gurman-2-3} conducted at 
filling factor $\nu_H=2/3$.
While the measured thermoelectric current,  plotted in Fig. S3 in the supplemental material of Ref.~\cite{gurman-2-3},
is very reminiscent of the thermopower potted in our Fig.~\ref{fig:TP} the 
red curve in Fig. 3c in Ref.~\cite{gurman-2-3} precisely corresponds to our Fig.~\ref{fig:PF} for the power factor,
including the observation that the measured quantity displays sharp dips at the centers of the conductance peaks.
Therefore, we expect that the power factor for SET transistors might be directly measurable by the method of
Ref.~\cite{gurman-2-3}.
\subsection{Even number of quasiparticle in the bulk}
\begin{figure}[htb]
\centering
\includegraphics[bb=10 15 570 390,clip,width=10cm]{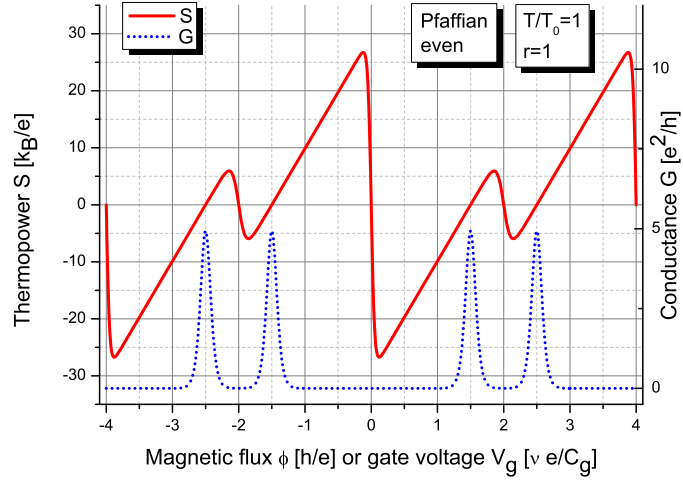}
\caption{Thermopower (left) and conductance (right) of the Pfaffian state with even number of bulk 
quasiparticles. \label{fig:TP2}}
\end{figure}

The case of the Pfaffian SET with even number of quasiparticles localized in the bulk is even more promising.
The disk partition function for this case is given by \cite{NPB2015}
\beq\label{Z-ev}
Z^{\mathrm{even}}(\tau,\zeta)=K_0(\t,2\z;8)\ch_0(\t') + K_4(\t,2\z;8)\ch_{1/2}(\t')
\eeq
where $\t'=(v_n/v_c) \t$, $K$ are the partition functions written in Eq.~(\ref{K}) and the neutral-sector partition functions 
are \cite{NPB2015}
\[
\ch_{0,1/2}(\t)= \frac{q^{-1/48}}{2}\left(\prod_{n=1}^\infty (1+q^{n-1/2}) \pm \prod_{n=1}^\infty (1-q^{n-1/2})\right),
\]
with $q$ defined as in Eq.~(\ref{K}).  Using the explicit partition function (\ref{Z-ev}) and substituting again in 
  Eqs.~(\ref{G}), (\ref{eps}), (\ref{N}),  and (\ref{H2})     
we compute the conductance and the thermopower for the Pfaffian state with even number of bulk quasiparticles 
and plot them for $T=T_0$  in Fig.~\ref{fig:TP2}.
Again the centers of the conductance peaks correspond to the zeros of the thermopower, however, in this case the 
conductance peaks  are grouped in pairs and the sawtooth curve of the thermopower is modulated in a corresponding way.
This modulation is due to the neutral sector of the corresponding CFT and its vanishes when the ratio $r=v_n/v_c$ 
becomes much smaller than 1 since then the role of the neutral part is decreased.

\begin{figure}[htb]
\centering
\includegraphics[bb=10 15 570 390,clip,width=10cm]{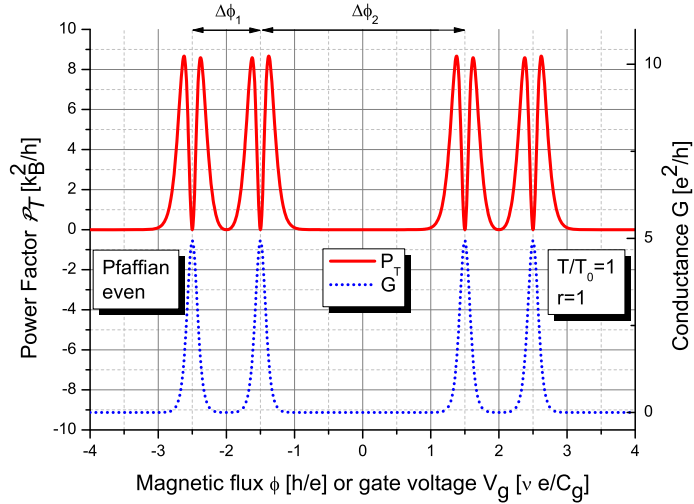}
\caption{Thermoelectric power factor (left) and conductance (right) of the Pfaffian state with even number of bulk 
quasiparticles. \label{fig:PF-even}}
\end{figure}

Next, using the computed thermopower for the Pfaffian FQH state with even 
number of bulk quasiparticles  and Eq.~(\ref{PF})  
we plot in Fig.~\ref{fig:PF-even} the power factor profile, together with the conductance \cite{NPB2015}
 for  $T=T_0$ with  $r=1$.
The modulation of power factor's dips could in general be used to estimate experimentally the ratio $r=v_n/v_c$ by measuring 
the two periods $\Delta\phi_1$ and $\Delta\phi_2$ between the dips because \cite{NPB2015} $\Delta\phi_1 =2-r$ and 
$\Delta\phi_2 =2+r$, see Fig.~\ref{fig:PF-even}.

\begin{figure}[htb]
\centering
\includegraphics[bb=25 70 550 770,clip,width=8cm]{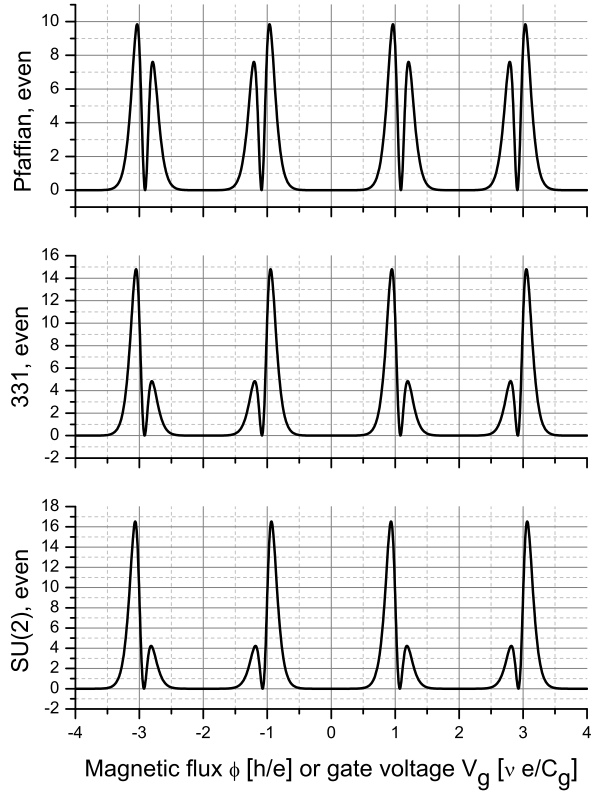}
\caption{Power factors for different candidates for $\nu_H=5/2$. \label{fig:PF1-6}}
\end{figure}

Finally we plot in Fig.~\ref{fig:PF1-6} the power factor profiles for several different candidate states, taken from Ref.~\cite{NPB2015}, 
one of which is expected  to describe the experimentally observed quantum Hall state at $\nu_H=5/2$ for  $T=T_0$ with small  
$r=1/6$ and even number of bulk quasiparticles.
This plot shows that even for smaller ratio $r$, when the neutral sector plays a subleading role the power factor is still sensitive
enough to distinguish between different states with identical zero-temperature electric characteristics.
At the same time, measuring the thermopower and power factor experimentally by the method of Ref.~\cite{gurman-2-3}, 
could allow us to estimate the ratio $r$ experimentally, by measuring the distances between the sharp dips of the power factor, 
since the spacing of the conductance peaks depends on $r$.
Furthermore, computing the power factors for these states at several different temperatures \cite{NPB2015}, 
e.g. $T/T_0=0.75, 1.00$ and $1.25$,
 and comparing them with  the power factors measured at these temperatures by the method of Ref.~\cite{gurman-2-3} will already 
allow us to distinguish between the candidates with non-Abelian anyons form those with Abelian anyons only
and eventually conclude from appropriate experiments whether non-Abelian anyons are indeed realized in the fractional quantum 
Hall states with filling factors $5/2$ and $12/5$.
\begin{acknowledgement}
This work has been partially supported by the Alexander von Humboldt Foundation under the Return Fellowship and 
Equipment Subsidies Programs and by the Bulgarian Science Fund under Contract No. DFNI-E 01/2 and  DFNI-T 02/6.
\end{acknowledgement}

\bibliographystyle{spphys}
\bibliography{CB,FQHE,my,QI,TQC,Z_k}
\end{document}